\documentclass[aps,prx,twocolumn,footinbib]{revtex4-2}
\usepackage{graphicx}
\usepackage{indentfirst}
\usepackage{physics}
\usepackage{braket}
\usepackage{float}
\usepackage{amsmath}
\usepackage{epstopdf}
\usepackage{footnote} 
\usepackage{CJK}
\usepackage{esint}
\usepackage{color}
\usepackage[T1]{fontenc}
\usepackage{subfigure}
\usepackage{amsfonts}
\usepackage{footmisc}
\usepackage{scrextend}
\usepackage{multirow}
\usepackage[hyperfootnotes=false]{hyperref}
\usepackage[acronym]{glossaries}

\usepackage[english]{babel}
\usepackage{url}
\usepackage{bm}
\usepackage{hyperref}
\definecolor{darkblue}{rgb}{0,0,0.5}
\hypersetup{
colorlinks=true,
linkcolor=black,
filecolor=blue,
citecolor=darkblue,  
urlcolor=black,
}

\newtheorem{theorem}{Theorem}

\newenvironment{proof}[1][Proof]{\noindent\textbf{#1.} }{\ \rule{0.5em}{0.5em}}

\newcommand{\1}{^{(1)}}

\def\be{\begin{equation}}
\def\ee{\end{equation}}
\def\ba{\begin{eqnarray}}
\def\ea{\end{eqnarray}}
\usepackage{bm}

\begin{document}
\title{Optimal noisy entanglement testing for ranging and communication}

\author{Pengcheng Liao$^1$}

\author{Quntao Zhuang$^{1,2}$}
\email{qzhuang@usc.edu}

\affiliation{
$^1$Ming Hsieh Department of Electrical and Computer Engineering, University of Southern California, Los
Angeles, California 90089, USA
\\
$^2$Department of Physics and Astronomy, University of Southern California, Los
Angeles, California 90089, USA
}

\begin{abstract}
Given a quantum system $S$ entangled with another system $I$, the entanglement testing problem arises, prompting the identification of the system $S$ within a set of $m \ge 2$ identical systems. This scenario serves as a model for the measurement task encountered in quantum ranging and entanglement-assisted communication [Phys. Rev. Lett. {\bf 126}, 240501, (2021)]. In this context, the optimal measurement approach typically involves joint measurements on all $m+1$ systems. However, we demonstrate that this is not the case when the subsystems containing system $S$ are subjected to entanglement-breaking noise.
Our approach utilizes the recently developed measurement technique of correlation-to-displacement conversion. We present a structured design for the entanglement testing measurement, implementable with local operations and classical communications (LOCC) on the $m+1$ systems. Furthermore, we prove that this measurement approach achieves optimality in terms of error probability asymptotically under noisy conditions.
When applied to quantum illumination, our measurement design enables optimal ranging in scenarios with low signal brightness and high levels of noise. Similarly, when applied to entanglement-assisted classical communication, the measurement design leads to a significant relative advantage in communication rates, particularly in scenarios with low signal brightness.

\end{abstract}
\maketitle


\section{Introduction}

Entanglement is one of the most intriguing phenomena in quantum mechanics. It manifests non-local correlation that remains even if two quantum systems are far apart, as evident from the Bell inequality tests~\cite{hensen2015loophole,hu2022high,storz2023loophole} and satellite-based entanglement distribution experiments~\cite{yin2017satellite}. As the `spooky
action at a distance' unveils itself, it becomes apparent that this phenomenon is beneficial for various information processing tasks such as sensing~\cite{giovannetti2006quantum,zhang2021distributed} and communication~\cite{bennett2002entanglement,hao2021entanglement}.

\begin{figure*}
    \centering
    \includegraphics[width=0.9\textwidth]{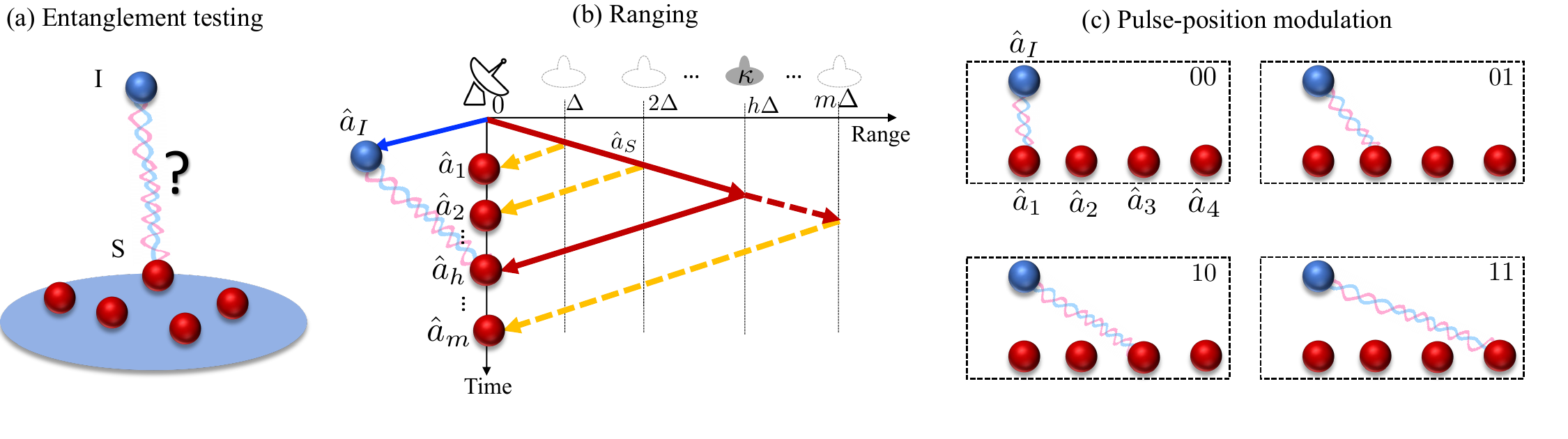}
    \caption{
    Schematic of (a) the problem of entanglement testing, and its application to (b) quantum ranging and (c) entanglement-assisted communication with pulse-position modulation. The blue ball depicts the reference photon, while the red balls depict the photons potentially entangled with the reference, the blue and red curvy lines indicate the original entangled signal-reference pair. 
    \label{fig:schematic}
    }
\end{figure*}

The subtle nature of entanglement makes measuring and manipulating it a challenging task.
We consider the example of entanglement testing (see Fig.~\ref{fig:schematic}a), where one needs to identify the signal system $S$---entangled with a given reference system $I$---among multiple identical systems. The entanglement testing problem models the measurement task encountered on the receiver's end during quantum ranging~\cite{zhuang2021quantum} and entanglement-assisted communication with pulse-position modulation~\cite{zhuang2021quantum,cariolaro2010theory}. 

In general, the optimal measurement approach for entanglement testing is expected to involve joint measurement of the entire $m+1$ systems, beyond current experimental capabilities when $m$ becomes large. Stepping back, designing any sub-optimal sequential measurement with local operations and classical communications (LOCC) is also non-trivial, especially in the optical domain relevant to ranging and communication.
This is because most available detectors are destructive—once the reference photon $I$ is measured, it is absorbed and therefore destroyed. The destructive nature of the measurement prevents a sequential scheme as a single reference $I$ can only support a two-photon measurement with one of the $m$ photons.
Indeed, previous experiments of quantum hypothesis testing in optical systems have all relied on sequential measurements, such as realizing the Dolinar receiver~\cite{cook2007optical,dolinar_processing_1973}, conditional nulling receiver~\cite{chen2012optical,dolinar1982near}, adaptive multi-state discrimination~\cite{becerra2013experimental,becerra2013implementation,ferdinand2017multi} and receivers empowered by machine learning~\cite{cui2022quantum}. These experiments consider quantum states without nonclassical correlation, while entanglement testing involves quantum correlations.

Such a unique quantum hypothesis testing problem not only reflects the fundamental conundrum of quantum measurement in manipulating entanglement, but is also relevant to various quantum sensing and communication applications. Here we list two examples---quantum ranging and entanglement-assisted communication (see Fig.~\ref{fig:schematic}bc). In time-of-flight quantum ranging~\cite{zhuang2021quantum}, the signal photon entangled with the reference idler may return to the detector among $m$ time bins corresponding to the different range of the reflecting target, among other noise photons. In entanglement-assisted communication with pulse-position modulation~\cite{zhuang2021quantum,cariolaro2010theory}, one encodes $\log_2 m$ bits of information by sending the signal pulse among $m$ positions (time bins), while the decoding side needs to decide which position (time bin) contains the photons entangled with the reference. 

Given the general challenging nature of the entanglement testing problem, we consider its noisy limit, where $m$ potential signals are contaminated with noise. Such a noisy limit of entanglement testing reduces to the testing of quantum discord~\cite{ollivier2001quantum}, which might admit easier optimal measurement strategy. 
At the same time, due to the reference idler being noiseless, the problem is not entirely classical---Ref.~\cite{zhuang2021quantum} finds that the error probability is suppressed by the initial entanglement beyond what is possible were the initial signal-idler correlation classical. However, the analyses is based on the quantum Chernoff bound~\cite{audenaert2007discriminating,Pirandola2008,nussbaum2011asymptotic,li2016discriminating}, while the measurement scheme for the entanglement testing problem is unknown.

In this work, we provide an explicit design of the entanglement testing measurement with LOCC and prove its optimality in the noisy limit when the signal brightness is low. We adopt a recently developed tool of correlation-to-displacement conversion~\cite{shi2022fulfilling,shi2023optimal} and combine it with a conditional nulling adaptive decision strategy. Via heterodyne detection on the signal system, one can convert the quantum correlation to coherent optical displacement, such that classical measurement strategies can follow up to resolve the quantum problem. Through asymptotic analyses, we offer a closed-form solution to the error probability, which aligns well with exact numerical simulations. In terms of the ranging application, an optimal six-decibel error-exponent advantage over classical schemes can be achieved by the proposed measurement. Interpreting the results to pulse-position-modulated entanglement-assisted communication, the proposed receiver achieves increasing advantage over the classical capacity without entanglement assistance~\cite{hausladen1996classical,schumacher1997sending,holevo1998capacity} as the signal brightness decreases, outperforming previous receivers that achieve only a constant advantage~\cite{shi2020practical}.

\section{Continuous-variable entanglement testing}
\label{sec:scenario}

As a single-photon state is susceptible to noise and loss, we consider the continuous-variable version of entanglement testing. In this version, the signal mode and idler mode are initially in the two-mode squeezed vacuum (TMSV) state, which is the multi-photon infinite-dimensional generalization of a Bell pair. The wave function of a pair of TMSV in the photon number basis is given by
\begin{equation}
\ket{\phi^{\text{TMSV}}}_{SI}=\sum_{n=0}^\infty \sqrt{\frac{N_{S}^{n}}{(N_{S}+1)^{n+1}}} \ket{n}_{S}\ket{n}_{I},
\label{eq:state_TMSV}
\end{equation}
where $\ket{n}$ is the number state defined by $\hat{a}^\dagger \hat{a}\ket{n}=n\ket{n}$, and $\hat{a}$ is the annihilation operator. Both the signal and the reference idler have a thermal reduced state with a mean photon number $N_{S}$.

We consider the noisy version of entanglement testing, where the signal goes through a thermal-loss channel $\mathcal{L}^{\kappa,N_B}$ described by the input-output relation of field annihilation operators
\begin{equation}
\hat{a}_h = \sqrt{\kappa} \hat{a}_S +\sqrt{1-\kappa}\hat{e}_h,   
\end{equation}
where $\kappa$ is the transmissivity and $\hat{e}_h$ is the thermal noise mode with a thermal photon number $N_B/(1-\kappa)$. It is straightforward to evaluate the mean photon number of the noisy return mode $\expval{\hat{a}_h^\dagger \hat{a}_h}=N_B+\kappa N_S\simeq N_B$, considering the limit $N_B\gg N_S$.

As shown in Fig.~\ref{fig:schematic}, the noisy version of continuous-variable entanglement testing considers the scenario where the noisy return is hidden among $m-1$ identical thermal modes $\hat{a}_\ell, 1\le \ell\neq h \le m$, each with a mean photon number $N_B$. The task of entanglement testing is to recognize $\hat{a}_h$ among the $m$ identical modes, given access to the perfectly stored idler $\hat{a}_I$. Mathematically, this is a quantum state hypothesis testing between
\begin{equation}
\label{eq:return-idler state}
   \hat{\rho}_h=\left(\otimes_{n \neq h} \hat{\sigma}_{\hat{a}_{n}}^{(B)}\right) \otimes \hat{\Sigma}_{\hat{a}_h \hat{a}_I}^{(T)},
\end{equation}
where $\hat{\sigma}^{(B)}$ denotes a thermal state with a mean photon number $N_B$, and $\hat{\Sigma}^{(T)}$ denotes the noisy TMSV when the signal goes through the thermal-loss channel. For more details on the characterization of the involved quantum states, refer to Ref.~\cite{zhuang2021quantum}.

When the noise $N_B$ is significantly high, even though the return $\hat{a}_h$ and idler $\hat{a}_I$ are no longer entangled, determining which pair was originally entangled remains a nontrivial task.

Despite the complexity arising from the involvement of a large number of systems ($m+1$), Ref.~\cite{zhuang2021quantum} demonstrates that the error probability in the weak signal and high noise limit ($N_B\gg1, N_S\ll1$) exponentially decays with the number of repeated probing $M$:
\begin{equation}
    P_{E,H} \sim \frac{m-1}{m}\exp\left[-\frac{2M\kappa N_S}{N_B}\right],
    \label{P_E_H_QCB}
\end{equation}
which is tight in the error exponent. Notably, this entanglement-assisted performance exhibits a 6-decibel error exponent advantage over scenarios where there is no entanglement between the signal $S$ and idler $I$ initially. For classical coherent state signals with the same total mean photon number $MN_S$, the error probability asymptotically follows:
\begin{align}
    P_{C,H}&\sim  \frac{m-1}{m}\exp\left[-\frac{2M\kappa N_S}{1+2N_B+2\sqrt{N_B\left(1+N_B\right)}}\right]
    \nonumber
    \\
    &\simeq \frac{m-1}{m}\exp\left[-\frac{M\kappa N_S}{2N_B}\right], \quad \text{when } N_B\gg1.
    \label{P_C_H_QCB}
\end{align}
This expression is also tight in the exponent.

Now, let's delve into the application of noisy entanglement testing in quantum ranging and entanglement-assisted communication. As depicted in Fig.~\ref{fig:schematic}b, the objective of ranging is to determine the distance of a reflecting target from the observer. For a required precision $\Delta$, the range can be discretized into $m$ bins of size $\Delta$, and the target's location in the $h$-th bin is unknown to the observer. To solve the ranging problem, the signal $\hat{a}_S$ is sent out, and returned signals are continuously collected at times $t_\ell= 2\ell\Delta/c $, where $c$ is the speed of light, and $1\le \ell \le m$. When the outgoing signal hits the target at distance $h\Delta$, it gets reflected back mixed with thermal noise. This process can be modeled by the thermal loss channel $\mathcal{L}^{\kappa,N_B}$, where $\kappa$ is the reflectivity of the target, and $N_B$ describes the noise. To enhance performance, quantum ranging stores an idler $\hat{a}_I$ for reference~\cite{zhuang2021quantum}. At the receiver side, the ranging task is reduced to entanglement testing—determining which mode among the $m$ return modes $\hat{a}_\ell, 1\le \ell \le m$ was originally entangled with the reference idler $\hat{a}_I$.

The scenario of entanglement-assisted communication with pulse-position modulation is akin to that of quantum ranging~\cite{zhuang2021quantum}. In Fig.~\ref{fig:schematic}c, the encoding involves two bits, resulting in four possible configurations of entanglement. Each configuration comprises a signal pulse $\hat{a}_S$ in a distinct time bin, entangled with the idler $\hat{a}_I$, which is pre-shared without noise to the receiver. After traversing a communication link modeled as the thermal loss channel $\mathcal{L}^{\kappa,N_B}$, the received modes $\hat{a}_1,\cdots \hat{a}_4$ all assume a thermal state with a mean photon number $N_B$. This is under the approximation that the source brightness $N_{S}$ is low. The decoder's objective is to determine which mode among $\hat{a}_1,\cdots \hat{a}_4$ was originally entangled with $\hat{a}_I$.

Having introduced the problem in a single-mode formalism, we can enhance the Signal-to-Noise Ratio (SNR) amid high noise by sending multiple modes in each pulse, denoted as $\hat{\boldsymbol{a}}_S \equiv \{\hat{{a}}_S^{(n)}\}_{n=1}^M$. Consequently, each return pulse $\hat{\boldsymbol{a}}_\ell \equiv\{\hat{{a}}_\ell^{(n)}\}_{n=1}^M$ contains $M$ modes, for $1\le \ell \le m$. Each $\hat{{a}}_S^{(n)}$ is entangled with an idler mode $\hat{{a}}_I^{(n)}$, resulting in the idler system $\hat{\boldsymbol{a}}_I \equiv \{\hat{{a}}_I^{(n)}\}_{n=1}^M$ containing $M$ modes as well. In both the ranging and communication scenarios, this configuration requires the time bin duration $T\equiv 2\Delta /c$ to satisfy $M\equiv TW\gg1$, where $W$ is the bandwidth of the signal.

\section{Measurement design}
As shown in Fig.~\ref{fig:receiver_design}, we propose a structured receiver design to solve the above entanglement testing problem. The receiver contains two steps. First, heterodyne detection is performed on the $m$ return systems to implement the correlation-to-displacement conversion~\cite{shi2022fulfilling,shi2023optimal}, which reduces the entanglement testing problem to the coherent state detection problem. Due to the specific structure of the coherent states, the second step then involves a conditional nulling detection to determine the system originally entangled with the reference idler.

\subsection{Converting Correlation to displacement}

\begin{figure}
    \centering
    \includegraphics[width=1.0\columnwidth]{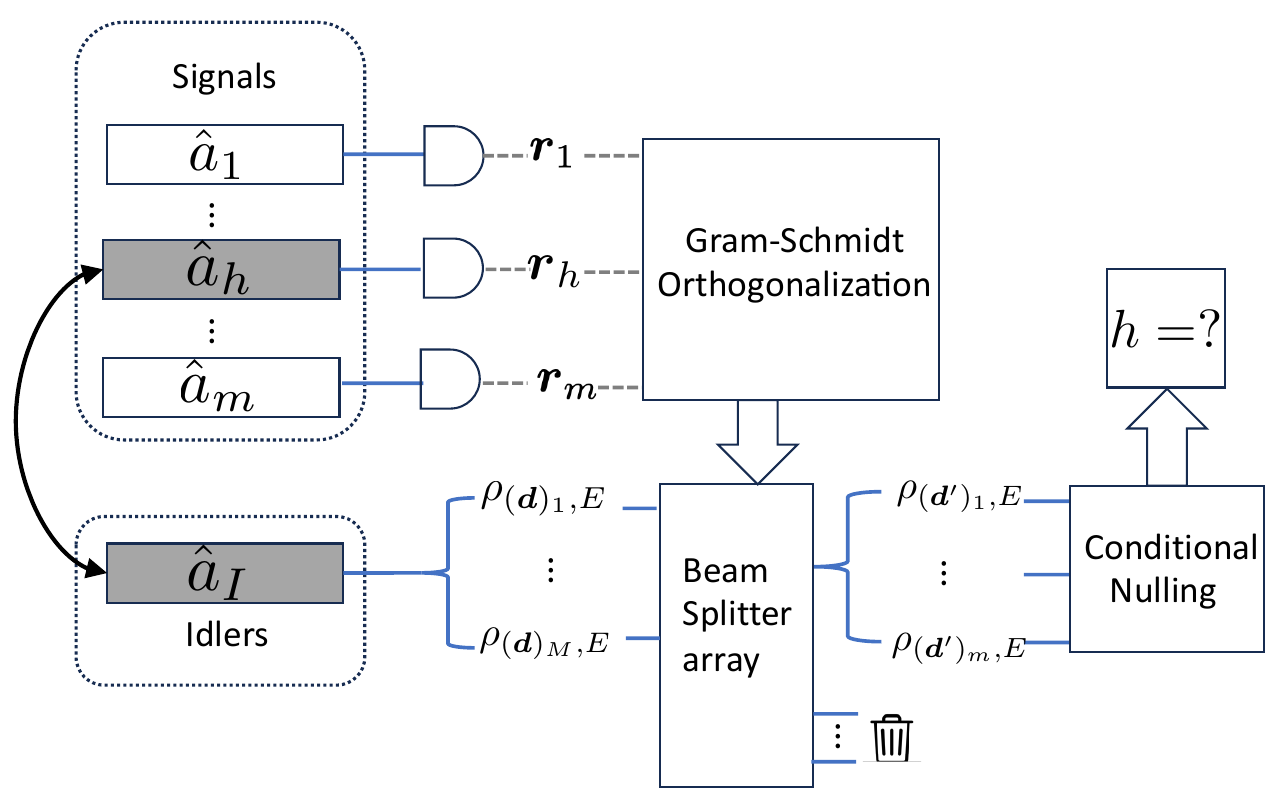}
    \caption{Illustration of the receiver design process. Initially, the measurement outcomes of the signal modes are employed to compute the weight matrix $W$ through Gram-Schmidt orthogonalization. This calculated matrix is then utilized to control the beam splitter. Subsequently, the conditional nulling strategy is applied to ascertain the accurate value of $h$.  }
    \label{fig:receiver_design}
\end{figure}

The first step of the receiver protocol adopts the correlation-to-displacement conversion, where one performs heterodyne measurement on all return pulses $\hat{\boldsymbol{a}}_1, \cdots, \hat{\boldsymbol{a}}_m$, leading to measurement results $\bm{r}_1,\cdots, \bm{r}_m$, with each $\bm{r}_i\in \mathbb{C}^M$ being a complex vector.
From Refs.~\cite{shi2022fulfilling,shi2023optimal}, each element of the vector $(\bm{r}_i)_j \in\mathbb{C} $ obeys a circularly-symmetric complex Gaussian distribution with the variance
\begin{equation}
v_\mathcal{M} = (N_B+\kappa N_S +1)/2  .
\end{equation}
Here we have adopted the notation $(\bm{x})_k$ to denote the $k$th component of the vector $\bm{x}$, where the extra bracket is added to avoid potential confusion with the system indices in the subscripts.

As indicated in Fig.~\ref{fig:receiver_design}, suppose the true hypothesis is $\hat{\rho}_h$ as specified in Eq.~\eqref{eq:return-idler state}, where the return pulse $\hat{\boldsymbol{a}}_h$ and the idler $\hat{\boldsymbol{a}}_I$ are correlated. Then, the idler has a remaining entanglement connection with $h$-th bin before the heterodyne measurement. Conditioned on the measurement,  
the idler mode $\hat{{a}}_I^{(n)}$ is in a displaced thermal state $\hat{\rho}_{(\bm{d})_n,E}$, with the mean 
\begin{align}
(\bm{d})_n = \frac{C_p}{2v_\mathcal{M}} (\bm{r}_h^*)_n \equiv \frac{\sqrt{\kappa N_S(N_S+1)}}{N_B+\kappa N_S +1} (\bm{r}_h^*)_n, 
\end{align}
and thermal mean photon number
\be 
E = \frac{N_S(N_B+1-\kappa)}{2v_\mathcal{M}}\equiv \frac{N_S(N_B+1-\kappa)}{N_B+\kappa N_S +1}.
\ee 
Here we define the notation $C_p\equiv \sqrt{\kappa N_S(N_S+1)}$ to simplify later calculations.
For simplicity, we also denote the conditional mean of all $M$ modes $\hat{\boldsymbol{a}}_I $ as the vector $\bm{d}$.

Conditioned on the measurement results $\bm{r}_1,\cdots, \bm{r}_m$, the correlation-to-displacement conversion then programs an array of beam splitters to interfere the $M$ idler modes and produces 
the first $m$ output modes
\be 
\hat{a}_I^{(n)\prime}=\frac{1}{c_n}\bm{r}_n^\prime \cdot \hat{\boldsymbol{a}}_I, 1\le n \le m,
\ee 
while the rest $M-m$ modes are discarded,
where $\{\bm{r}_n^{\prime}\}_{n=1}^m$ is obtained by Gram-Schmidt orthogonalization of the measurement result vectors $\{\bm{r}_n\in \mathbb{C}^{M} \}_{n=1}^m $ and each $c_n = |\bm{r}_n^{\prime}|$ is the normalization constant. At this point, the correlation-to-displacement conversion step is complete. Below we provide analytical insights to the conversion.

To facilitate our analyses, we define the $M\times M$ weight matrix to represent the beamsplitter transform
\begin{equation}
W = 
\begin{pmatrix}
\bm{r}_1^{\prime T}/c_1 \\
\vdots \\
\bm{r}_m^{\prime T}/c_m\\
\vdots 
\end{pmatrix}.
\end{equation}
Here we only explicitly express the first $m$ rows. This choice stems from our focus on the initial $m$ modes, with the remaining $M-m$ modes deemed irrelevant and subsequently discarded. The omitted $M-m$ rows of the weight matrix can be chosen arbitrarily subject to the orthogonality constraint. With the notation defined, we can write the mean of the $m$ idler modes after the beamsplitter array as
\begin{align}
\bm{d}'
= W \cdot \bm{d}
= \frac{C_p}{2v_\mathcal{M}} 
\begin{pmatrix}
\bm{r}_1^{\prime T} \cdot \bm{r}_h^*/c_1 \\
\vdots  \\
\bm{r}_m^{\prime T} \cdot \bm{r}_h^*/c_m
\end{pmatrix}.
\end{align}
Note that the thermal noise characteristics among the idler modes are not changed as the thermal noise is independent and identical (iid) among all modes. Consequently, after the entire correlation-to-displacement conversion the idler mode $\hat{{a}}_I^{(n)}$ is in a displaced thermal state $\hat{\rho}_{(\bm{d}^\prime)_n,E}$.

As different $\bm{r}_i$'s are independently sampled from high-dimensional ($M\gg 1$) space, they are almost orthogonal, $\bm{r}_i^T\cdot \bm{r}_j^* / |\bm{r}_i||\bm{r}_j| \approx 0$. More specifically, for any constant $a>0$, the following probability bounds hold:
\begin{subequations}
\label{Pr_bound}
\begin{align}
\operatorname{Pr} \left( |\operatorname{Re} (\bm{r}_i^T\cdot \bm{r}_j^*)| / |\bm{r}_i||\bm{r}_j| >a \right) \le \frac{1}{2a^2 M},\\
\operatorname{Pr} \left( |\operatorname{Im} (\bm{r}_i^T\cdot \bm{r}_j^*)| / |\bm{r}_i||\bm{r}_j| >a \right) \le \frac{1}{2a^2 M}.
\end{align}  
\end{subequations}
See Appendix~\ref{sec:ortho_aprox} for a proof.
As a result, the Gram-Schmidt orthogonalization does not change the vectors a lot: $\bm{r}_i^{\prime}\approx \bm{r}_i$. 
An experimental realization of such a beamsplitter transformation can potentially be via extending the mode selector in Ref.~\cite{shi2023optimal}, with programmable mode coupler, phase modulation and dispersion, or sum-frequency-generation.

With the above approximation, the mean of the idler states become
\begin{align}
\bm{d}'
\approx
\frac{C_p}{2v_\mathcal{M}} 
\begin{pmatrix}
\bm{r}_1^{ T} \cdot \bm{r}_h^*/c_1 \\
\vdots  \\
\bm{r}_m^{ T} \cdot \bm{r}_h^*/c_m
\end{pmatrix},
\end{align}
in which 
\begin{align}
\label{eq:alpha1}
(\bm{d}^\prime)_h\equiv \alpha_1 
 \approx C_p \sqrt{\frac{M}{2v_\mathcal{M}}} 
 =\sqrt{\frac{M\kappa N_S(N_S+1)}{(N_B+\kappa N_S +1)}},
\end{align}
while all other entries are $ (W \cdot \bm{d})_i  \equiv \alpha_0 \approx 0$. At the same time, the statistical fluctuation of mean due to the random outcome is small compared to thermal noise $E$, when $N_B\gg1$ (see Appendix~\ref{sec:ortho_aprox}). This means that we can use the above approximation of $\alpha_1,\alpha_0$ in our asymptotic analyses. At this point, the correlation-to-displacement conversion module has transformed the original entanglement testing problem to a coherent state testing problem, where there is a non-zero mean pulse among $m-1$ zero-mean pulses, in presence of common noise of $E$ on all modes.

\subsection{Coherent-state detection: Conditional-nulling}

Now we design the measurement for coherent state hypothesis testing, which completes the measurement design for noisy entanglement testing.
As the problem is state hypothesis testing between pulse-position modulated coherent states, we directly adopt the generalized conditional-nulling receiver~\cite{dolinar1982near,chen2012optical,zhuang2020entanglement}. It involves optical displacements and on-off photon detection measurements, with the adaptive strategy as described below:
\begin{enumerate}

\item  Choose variable $i = 1$, then perform displacement $D(-\alpha_1)$ on the $i$-th mode;

\item Perform binary state discrimination on the $i$-th mode via photon counting: 
if the count is non-zero, discard the $i$-th mode, and repeat step 1 with the index increased to $i+1$; 
if the count is zero, go to the next step;

\item  Perform binary state discrimination via photon counting sequentially on the remaining modes $j=i+1,\cdots m$. 
If the $j$-th mode count is non-zero, conclude with the decision $\tilde{h}=j$. If all the remaining counts are zero, then conclude with the decision $\tilde{h}=i$.
\end{enumerate}
To understand the performance, we consider the two types of errors happening in the binary state discrimination sub-problem: false alarm error, where one mistakes zero mean as non-zero and we denote the error probability as $p_1$; false negative error, where one mistakes non-zero mean as zero and we denote the error probability as $p_2$. We can gain insights into the problem by first considering the $M\gg1$ limit, where the statistical fluctuations in $\alpha_1$ and $\alpha_0$ are negligible and therefore
\begin{subequations}
\label{eq:p1p2}
\begin{align}
&p_1 =     1- P_{\alpha_0, E}(0)=1-e^{-\frac{|\alpha_0|^2}{E+1}} \frac{1}{1+E}, 
\\
&p_2= P_{\alpha_1 , E}(0)=e^{-\frac{|\alpha_1|^2}{E+1}} \frac{1}{1+E},
\end{align}
\end{subequations}
where $\alpha_0 \approx 0$, $\alpha_1 
 \approx C_p \sqrt{\frac{M}{2v_\mathcal{M}}}$ as specified in Eq.~\eqref{eq:alpha1}, and $P_{\alpha, N_B}(\cdot)$ is the number distribution of displaced thermal state with mean $\alpha$ and thermal noise $N_B$.

\section{Performance analysis}
\subsection{Error probability}
We begin the performance analysis with the error probability in noisy entanglement testing, which has a direct implication in entanglement-assisted ranging. We combine both analytical approach and Monte-Carlo simulations in our analyses. To enable analytical solution, we take the $M\gg1$ approximation such that the statistical fluctuation is not important and two types of error are well characterized by Eqs.~\eqref{eq:p1p2}. In this case, the overall error probability $P_m$ is a function of the error probabilities of the binary state discrimination sub-task, false-alarm probability $p_1$ and false-negative probability $p_2$. However, due to the thermal noise, unambiguous information is impossible and the derivation in Ref.~\cite{zhuang2020entanglement} does not apply. Towards this end, we derive the iterative formula as the following.
\begin{theorem}
\label{theorem:iteration}
The error probability $P_m$ for the conditional nulling receiver with $m$ equal-prior hypotheses can be obtained via iterating 
\begin{align}
\label{eq:conditional_nulling_P}
 1-P_n = &\frac{(1-p_1)^n}{n}   
 \nonumber
 \\
 &+ \frac{n-1}{n} \left[\left(1-p_2\right)\left(1-P_{n-1}\right)+p_2 Q_{n-1} \right]
\end{align}
for $n=2, \cdots, m$ with the initial condition $P_1=0$,
where $p_1,p_2$ are the binary state discrimination error probabilities and the function
\begin{equation}
Q_{n} = 
\begin{cases}
1-p_2, \mbox{if } p_1=0,\\
{\left(1-p_2\right)\left[1-\left(1-p_1\right)^n\right]}/{\left(np_1\right)}, \mbox{otherwise}.
\end{cases}
\end{equation}
\end{theorem}
We present the proof in Appendix~\ref{app:proof}.

As a sanity check, we consider the extreme case where $N_S\ll1$ while $M\kappa N_S/N_B$ remains finite. At this limit, we can approximate $E=0$, $\alpha_0=0$ and $\alpha_1=\sqrt{ \frac{M\kappa N_S}{N_B}}$ in Eqs.~\eqref{eq:p1p2} and the displaced thermal states become coherent states, thus $p_1=0$ and $p_2=e^{-|\alpha_1|^2}$. This represents the ideal classical conditional nulling receiver scenario and the recursion relation in Eq.~\eqref{eq:conditional_nulling_P} becomes
\begin{equation}
 P_n = \frac{n-1}{n} \left( (1-e^{-|\alpha_1|^2} )P_{n-1}+ e^{-2|\alpha_1|^2} \right),
\end{equation}
which admits the closed-form solution
\begin{align}
P_{\rm cn,ideal}&\equiv P_m=\frac{1}{m}\left[me^{-|\alpha_1|^2}+\left(1-e^{-|\alpha_1|^2}\right)^m-1\right]
\label{asymptotic_PE_exact}
\\
&\approx \frac{1}{2}(m-1)e^{-2|\alpha_1|^2}=\frac{1}{2}(m-1)e^{-2M\kappa N_S/N_B}.
\label{asymptotic_PE}
\end{align}

\begin{figure}
    \centering
    \includegraphics[width=1.0\columnwidth]{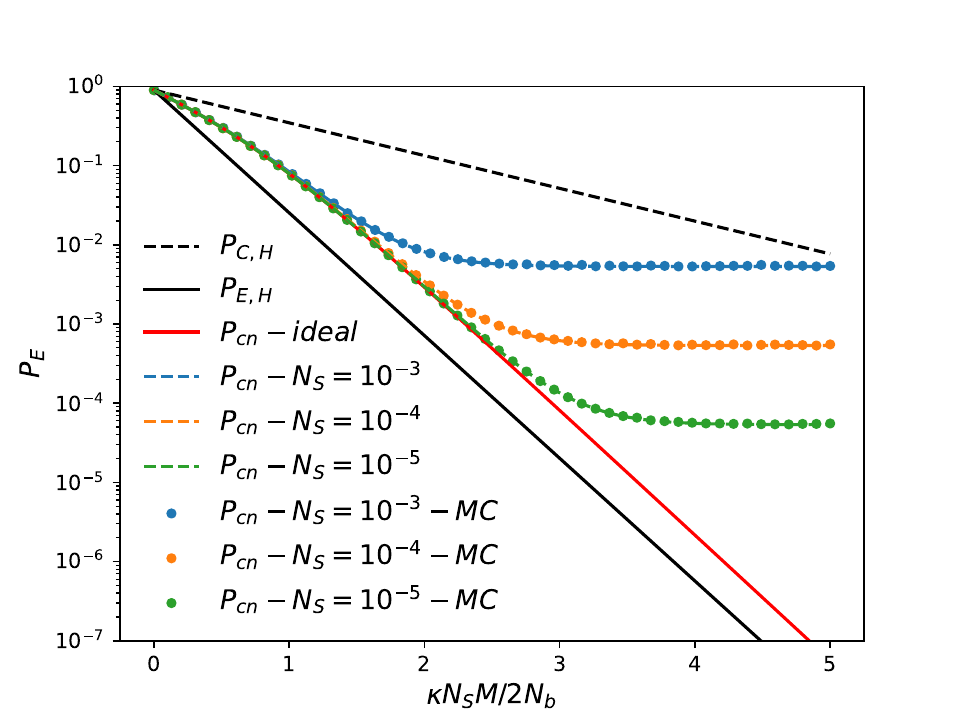}
    \caption{Error probability vs SNR, where $\kappa=0.1,N_b=10,m=10$, and $M\in[10/N_S,10^3/N_S]$. $P_{\rm cn}$ is the conditional nulling error calculated using the recursion relation Eq.~\eqref{eq:conditional_nulling_P} and Eq.~\eqref{eq:p1p2};  $\alpha_0=0$ and $\alpha_1$ is from Eq.(\ref{eq:alpha1}) in the dashed line; $\alpha_0$ and $\alpha_1$ are calculated from Monte Carlo simulation in the dotted line. The ideal-cn line corresponds to the case where the noise is zero. $P_{C,H}$ an $P_{E,H}$ are the classical and entanglement-assisted Helstrom limit, respectively.  }
    \label{fig:Pe_vs_snr}
\end{figure}


Comparing the receiver asymptotic performance in Eq.~\eqref{asymptotic_PE} and the ultimate quantum limit in Eq.~\eqref{P_E_H_QCB}, we prove that the proposed correlation-to-displacement based receiver achieves the optimal performance in noisy entanglement testing at the $N_S\ll1$ and $N_B\gg1$ limit. This is also verified numerically in Fig.~\ref{fig:Pe_vs_snr}, where Eq.~\eqref{P_E_H_QCB} (black solid) is just a constant factor better than the proposed receiver's asymptotic performance (Eq.~\eqref{asymptotic_PE_exact}, red solid).

At finite $N_S,N_B$, we evaluate the receiver performance via both exact recursion in Eq.~\eqref{eq:conditional_nulling_P}, as shown by the dashed lines with the color indicating different $N_S$ values. In the recursion, we have also taken the approximation that $\alpha_1  
 \approx \sqrt{\frac{M\kappa N_S(N_S+1)}{(N_B+\kappa N_S +1)}}$ and $\alpha_0\approx 0$, assuming $M\gg1$ making statistical fluctuations negligible. The results are also verified by exact Monte-Carlo simulation (dots) based on implementing the strategy and simulating the exact measurement statistics, including the heterodyne outcome $\bm{r}_1,\cdots, \bm{r}_m$, the corresponding Gram-Schmidt orthogonalization, and the randomness in $\alpha_0,\alpha_1$. 
This alignment of the dots and the dashed lines suggests that the approximation made in Eq.~\eqref{eq:alpha1} assuming $M\gg1$ is reasonable. Overall, the results at finite $N_S$ shows that the proposed receiver's error probability is bounded by $p_1$ and $p_2$, approaching a constant as the SNR becomes excessively large, with the lower bound decreasing in proportion to $N_S$. Therefore, the receiver's optimality only holds in the $N_S\ll1$ limit and at finite $N_S$ the optimal measurement for entanglement testing problem is still unknown with the presence of noise. It is likely that a joint detection instead of the conditional nulling type of LOCC strategy is needed.

\subsection{EA communication rates}
Now we translate the error probability performance of noisy entanglement testing to the application scenario of entanglement-assisted communication. As we explained in Sec.~\ref{sec:scenario}, in a pulse-position modulated communication protocol, one sends a pulse among $m$ time bins through the thermal-loss channel. To enhance the communication quality, one can also adopt a repetition encoding under the PPM encoding, with $M$ modes in each time bin. On the receiver side, suppose the error probability of the hypothesis testing is $P$, the information rate per mode can be evaluated from standard formula as $R_{m,M}(P)=I\left(P\right)/Mm$, where the mutual information
\begin{equation} 
I\left(p\right)=\log_2\left(m\right)+\left[\left(1-p\right)\log_2\left(1-p\right)+p\log_2\left(\frac{p}{m-1}\right) \right].
\end{equation} 
From the quantum limit $P_{E,H}$ in Eq.~\eqref{P_E_H_QCB}, we evaluate the rate allowed by entanglement-assisted PPM encoding as $R_{H}=\max_{m,M}R_{m,M}(P_{E,H})$; For the receiver performance, we use the iterative formula $P_m$ obtained from Eq.~\eqref{eq:conditional_nulling_P} under the $M\gg1$ limit to obtain $R_{\rm cn}=\max_{m,M}R_{m,M}(P_m)$. 

To benchmark the communication rates, we consider the ultimate rate limits of reliable communication over the quantum channel---the channel capacity.
Without entanglement-assistance, the achievable rates are bounded by the Holevo-Schumacher-Westmoreland (HSW) classical capacity~\cite{hausladen1996classical,schumacher1997sending,holevo1998capacity}, solvable for the thermal-loss channel as~\cite{giovannetti2014ultimate}
\be 
\label{eq:R_CA}
C(\mathcal{L}^{\kappa,N_B})=g\left(\kappa n_S+N_B\right)-g\left(N_B\right).
\ee 
where $g(n)=(n+1)\log_2(n+1)-n\log_2 n$ is the entropy of a thermal state with mean photon number $n$ and $n_S = N_S/m$ is photon number per mode when $Mm$ modes are considered. With entanglement, the EA classical capacity is increased to~\cite{bennett2002entanglement}
\be
C_E(\mathcal{L}^{\kappa,N_B})=g(n_S)+g(n_S^\prime)-g(A_+)-g(A_-),
\label{CE_formula}
\ee 
where 
$A_\pm=(D-1\pm(n_S^\prime-n_S))/2$, $n_S^\prime=\kappa n_S+N_B$ and $D=\sqrt{(n_S+n_S^\prime+1)^2-4\kappa n_S(n_S+1)}$. 
In the limit of $n_S\ll1, N_B\gg1$, we have
\begin{align}
&C(\mathcal{L}^{\kappa,N_B})\simeq \frac{\kappa n_S}{\ln(2)N_B},
C_E(\mathcal{L}^{\kappa,N_B}) \simeq \frac{\kappa n_S\ln(n_S)}{\ln(2)N_B},
\end{align}
showing $C_E/C\sim \ln (n_S)$ diverging as the brightness of the signal decreases to zero.

\begin{figure}
    \centering
    \includegraphics[width=1.0\columnwidth]{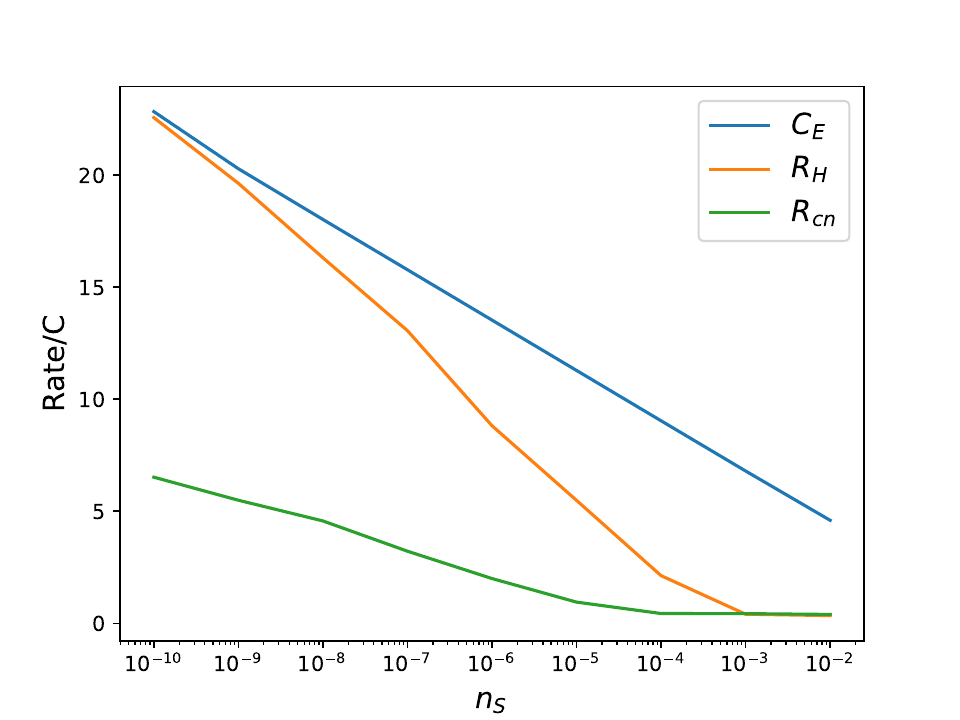}
    \caption{ $\text{Rate}/C$ versus $n_S$ for $N_b = 20$ and $\kappa = 0.1$. The values of $C$ and $C_{E}$ are determined using Eq. (\ref{eq:R_CA}) and Eq. (\ref{CE_formula}), respectively. Additionally, $R_{H}$ and $R_{\rm cn}$ are computed as $R_{L,M} = I\left(P\right)/Mm$ based on the ultimate quantum limit given by Eq. (\ref{P_E_H_QCB}) and the condition for nulling error probability provided in Eq. (\ref{eq:conditional_nulling_P}). The optimization is performed over the parameters $M$ and $m$ to maximize the corresponding values.}

    \label{fig:rate_vs_snr}
\end{figure}

The numerical results presented in Fig.~\ref{fig:rate_vs_snr} reveal interesting insights. 
As shown in Ref.~\cite{zhuang2021quantum}, in the limit of small $n_S$, the quantum limit of PPM entanglement-assisted communication $R_{H}$ approaches the entanglement-assisted capacity $C_E$ asymptotically, but a growing gap emerges as $n_S$ increases. On the other hand, there is a widening disparity between $R_{H}$ and rate offered by the proposed receiver $R_{cn}$ as $n_S$ decreases. 
This indicates that the correlation-to-displacement conversion condition-nulling receiver is unable to fully leverage the advantages of EA classical capacity in scenarios with lower signal strength. 
However, as the signal brightness diminishes, the condition-nulling receiver exhibits an expanding advantage over classical capacity, surpassing the performance of previous practical receivers~\cite{shi2020practical} that only maintain a constant advantage.

\section{Discussion and conclusion}

In this study, we introduce a structured receiver design aimed at addressing the continuous-variable entanglement testing problem in the presence of noise. In the asymptotic regime of low signal brightness, the receiver demonstrates optimal error-probability performance. However, in regions characterized by finite brightness and lower levels of noise, determining the optimal receiver design remains an open challenge.

Regarding entanglement-assisted communication, our proposed receiver efficiently decodes pulse-position modulation, exhibiting a rate advantage over classical capacity. Notably, this advantage increases as the signal brightness decreases. It is important to highlight that, while our receiver outperforms classical capacity, it does not yet reach the entanglement-assisted capacity. The task of devising an optimal receiver to fully saturate the entanglement-assisted capacity in decoding pulse-position-modulated signals remains an open problem.

\begin{acknowledgements}
The project is supported by Cisco Systems Inc., National Science Foundation OMA-2326746,
National Science Foundation Grant No. 2330310,
National Science Foundation CAREER Award CCF-2240641, 
Office of Naval Research Grant No. N00014-23-1-2296 and
National Science Foundation Engineering Research Center for Quantum Networks Grant No. 1941583. QZ acknowledges discussions with Jeffrey H. Shapiro.
\end{acknowledgements}

\appendix


\begin{figure*}
    \centering
    \includegraphics[width=0.9\textwidth]{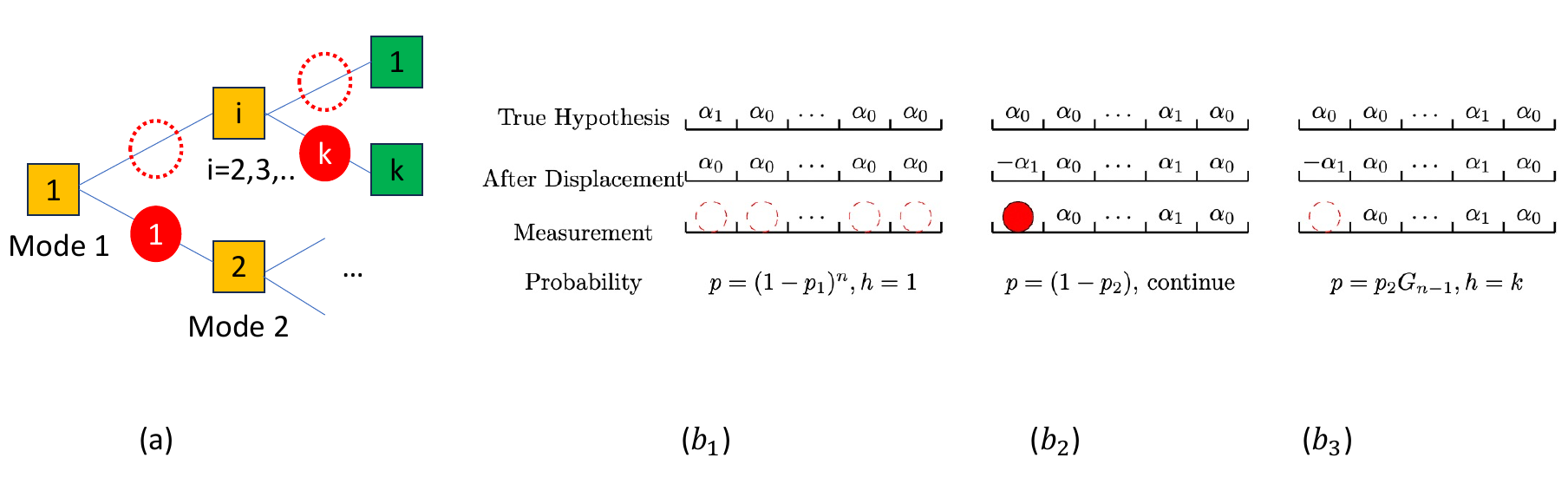}
    \caption{Potential trajectories and outcomes in the conditional nulling strategy. 
    (a) The procedure initializes setting $i=1$, and subsequent to the displacement $D(-\alpha_1)$, conducts photon counting in the initial mode. If the outcome is zero (depicted as an empty circle), one proceeds to photon counting sequentially for the remaining modes. In the event of a non-zero photon count in the $k$-th mode, the procedure ends with the decision that $\tilde{h}= k$; however, if all the remaining modes exhibit a photon count of zero, the procedure ends with the decision $\tilde{h}= 1$. Conversely, if the photon count in the first mode is non-zero (indicated by a solid circle), one eliminates the first mode and iterate the process starting from $i=2$.
    ($b_1$) Assuming the true hypothesis is $h=1$ and conditional nulling initiates with $i=1$, after the displacement $D(-\alpha_1)$ in the initial mode, the means of all modes become $\alpha_0\approx 0$. Achieving the accurate decision necessitates photon counts of 0 for all $n$ modes, illustrated by the dashed-line circle. The associated probability is denoted by $p = (1-p_1)^n$. This is denoted by the trajectory $1\to i\to 1$ in (a).
    ($b_2$) Assuming the true hypothesis is $\ell=k$ and conditional nulling commences with $i=1$, the mean of the first mode following the displacement is $-\alpha_1$. If the photon count in the first mode is non-zero, the first mode is discarded, and this iterative process is then applied to the remaining modes, as illustrated by the path $1\to 2\to \cdots$ in (a).
    ($b_3$) Under the same hypothesis as ($b_2$), if the photon count in the first mode is zero after the displacement, the subsequent step involves photon counting for all the remaining modes. Here, $G_{n-1}$ represents the probability of successfully identifying $\tilde{h} = k$, depicted by the path $1\to i\to k$ in (a).
    }
    \label{fig:cnpath}
\end{figure*}


\section{Orthogonality approximation}
\label{sec:ortho_aprox}

Let's assume $(\bm{r}_i)_j = a_{i j}+ib_{i j}$, thus both $a_{ij}$ and $b_{ij}$ obey a zero-mean Gaussian distribution with variance $v_\mathcal{M}$, $\mathcal{N}(0, v_\mathcal{M})$, leading to
\begin{align}
|\bm{r}_i| =&  \sqrt{\sum_{j=1}^M |(\bm{r}_i)_j|^2} 
= \sqrt{\sum_{j=1}^M (a_{ij}^2 + b_{ij}^2 )}  \\
=& \sqrt{v_\mathcal{M}} \sqrt{\sum_{j=1}^M \left( \frac{a_{ij}}{\sqrt{v_\mathcal{M}}} \right)^2 + \left( \frac{b_{ij}}{\sqrt{v_\mathcal{M}}} \right)^2}\\
=& \sqrt{v_\mathcal{M}} C,
\end{align}
where $C$ obeys the $\chi$-distribution of degree of freedom $2M$. When $M\gg 1$, the mean $\mathbb{E}(C) \approx \sqrt{2M}$ while the variance $\mathbb{VAR}(C) \approx 1/2$, so 
\begin{equation}
\mathbb{E}(|\bm{r}_i|)  \approx \sqrt{2M v_\mathcal{M}}, \quad
\mathbb{VAR}({|\bm{r}_i|}) \approx  v_\mathcal{M}/2.
\end{equation}
On the other hand, 
\begin{align}
\frac{\bm{r}_i^T\cdot \bm{r}_j^*}{|\bm{r}_i||\bm{r}_j|}= 
\frac{\sum_{l=1}^M (a_{il}+ib_{il})(a_{jl}-ib_{jl})}{|\bm{r}_i||\bm{r}_j| } \\
=\frac{\sum_{l=1}^M (a_{il}a_{jl}+b_{il}b_{jl})+i(a_{jl}b_{il}-a_{il}b_{jl})}{|\bm{r}_i||\bm{r}_j| }. 
\end{align}
By the law of large numbers, when $M\gg1$, $\sum_{l=1}^M a_{il}a_{jl}$ approximately obeys a Gaussian distribution $\mathcal{N}(0, Mv_\mathcal{M}^2)$, thus we can have
\begin{align}
\frac{|\bm{r}_i|}{\sqrt{2M v_\mathcal{M}}}\sim \mathcal{\chi}_{2M}(\mu=1,\sigma^2 =\frac{1}{4M} )    \\
\frac{\sum_{l=1}^M a_{il}a_{jl}}{ 2M v_\mathcal{M}}\sim \mathcal{N}(\mu=0,\sigma^2 =\frac{1}{4M} )    .
\end{align}
Therefore, when $M\gg1$, $\frac{|\bm{r}_i|}{\sqrt{2M v_\mathcal{M}}}\approx 1$ and $\frac{\sum_{l=1}^M a_{il}a_{jl}}{ 2M v_\mathcal{M}}\approx 0$. On the other hand,
${\bm{r}_i^T\cdot \bm{r}_j^*}/{|\bm{r}_i||\bm{r}_j|}$ can be re-writen as 
\begin{align}
&\frac{\sum_{l=1}^M (\frac{a_{il}a_{jl}}{2M v_\mathcal{M}}+ \frac{b_{il}b_{jl}}{2M v_\mathcal{M}} )
+i(\frac{a_{jl}b_{il}}{2M v_\mathcal{M}}-\frac{a_{il}b_{jl}}{2M v_\mathcal{M}})  }
{ \frac{|\bm{r}_i|}{\sqrt{2M v_\mathcal{M}}} \frac{|\bm{r}_j|}{\sqrt{2M v_\mathcal{M}}}  }     \\
\approx & \sum_{l=1}^M (\frac{a_{il}a_{jl}}{2M v_\mathcal{M}}+ \frac{b_{il}b_{jl}}{2M v_\mathcal{M}} )
+i(\frac{a_{jl}b_{il}}{2M v_\mathcal{M}}-\frac{a_{il}b_{jl}}{2M v_\mathcal{M}}) \\
=& R+i I, 
\end{align}
in which both $R$ and $I$ obey the distribution $\mathcal{N}(\mu=0,\sigma^2 =\frac{1}{2M} )$.
By Chebyshev's inequality $\operatorname{Pr} (|R-\mu|\ge k\sigma )\le \frac{1}{k^2}$ and setting $k = a/\sigma$, we have
\be
\operatorname{Pr} (|R|\ge a)\le \frac{\sigma^2}{a^2} = \frac{1}{2Ma^2}.
\ee
The analysis is the same for the imaginary part, thus we obtain Ineq.~\eqref{Pr_bound}.

Next, after the beamsplitter transformation,
the major non-zero mean of the mode is
\begin{align}
 &(W \cdot \bm{d})_h = \alpha_1= \frac{C_p}{2v_\mathcal{M}} |\bm{r}_h| \\
 \sim &\mathcal{\chi}_{2M}(\mu = C_p\sqrt{\frac{M}{2v_\mathcal{M}}},\sigma^2 =\frac{C_p^2}{8v_\mathcal{M}} ). 
\end{align}
More explicitly, the mean and the variance are
\begin{align}
\mu = \sqrt{\frac{\kappa M N_S(N_S+1)}{N_B+\kappa N_S+1}},    \\
\sigma^2 = \frac{\kappa N_S(N_S+1)}{4(N_B+\kappa N_S+1)}.
\end{align}

Using approximation $|\bm{r}_i|/\sqrt{2Mv_\mathcal{M}}\approx1$ when $M\gg1$,  the mode mean close to zero is
\begin{align}
&(W \cdot \bm{d})_i = \alpha_0 =\frac{C_p}{2v_\mathcal{M}}  \frac{\bm{r}_i^T\cdot \bm{r}_j^*}{|\bm{r}_i|} \\
=& \frac{C_p}{2v_\mathcal{M}} \sqrt{2Mv_\mathcal{M}}  \frac{\bm{r}_i^T\cdot \bm{r}_j^*/ 2Mv_\mathcal{M} }{|\bm{r}_i|/\sqrt{2Mv_\mathcal{M}}}\\
\approx & C_p \sqrt{\frac{M}{2v_\mathcal{M}}}  (R+iI) = R^\prime + iI^\prime,
\end{align}
in which 
\begin{align}
R^\prime, I^\prime \sim    \mathcal{N}(\mu=0,\sigma^2 =\frac{C_p^2}{4v_\mathcal{M}} ).
\end{align}

The variance of both $\alpha_1$ and $\alpha_0$ are proportional to $C_p^2/2v_\mathcal{M}$. 
Up to this point, we have only taken the $M\gg1$ limit and obtain asymptotic results that hold for any values of $N_S,N_B$. Now we further consider the $N_B\gg 1$ limit, both variances of $\alpha_0$ and $\alpha_1$ are very small compared to thermal noise $E$, so the following approximation holds,
\begin{align}
\alpha_1 \approx    C_p\sqrt{\frac{M}{2v_\mathcal{M}}}, \quad
\alpha_0 \approx 0,
\end{align}
as long as one includes the thermal noise $E$ in the quantum state.


\section{Proof of Theorem~\ref{theorem:iteration}} 
\label{app:proof}

\begin{proof}
At any step $i$, there are three scenarios in which the decision is correct;
\begin{enumerate}
\item  When the true hypothesis $h=i$, and the state discrimination gives the correct answer. This happens with the probability $\frac{(1-p_1)^n}{n}$ (path $1\to i\to 1$ in Fig.~\ref{fig:cnpath});

\item  When the true hypothesis $h\ne i$, and the state discrimination in step 2 gives non-zero mean correctly. This happens with the probability $\frac{n-1}{n} (1-p_2)(1-P_{n-1})$ (path $1\to 2\to \cdots$ in Fig.~\ref{fig:cnpath});

\item  When the true hypothesis $h=j\ne i$, and the state discrimination in step 2 gives zero mean incorrectly and finds the true $j$ in the remaining $n-1$ modes. This happens with the probability $\frac{n-1}{n} p_2 Q_{n-1} $ (path $1\to i\to k$ in Fig.~\ref{fig:cnpath}).
\end{enumerate} 
In summary,
\begin{align}
 1-P_n = \frac{(1-p_1)^n}{n}   + \frac{n-1}{n} (1-p_2)(1-P_{n-1}) \\
 + \frac{n-1}{n} p_2 Q_{n-1},
\end{align}
where $Q_{n-1}$ is the probability of successfully identifying the correct mode using one-by-one method. We have
\begin{equation}
 Q_{n-1} = \frac{1}{n-1} (1-p_2) +\frac{n-2}{n-1}(1-p_1) Q_{n-2}, 
\end{equation}
so 
\begin{equation}
Q_{n} = 
\begin{cases}
1-p_2, \text{if}\quad p_1=0\\
\frac{(1-p_2)(1-(1-p_1)^n)}{np_1}, \text{otherwise}.
\end{cases}
\end{equation}
\end{proof}

%

\end{document}